\documentclass[aps,twocolumn,notitlepage,showpacs,superscriptaddress,10pt,footinbib,floatfix]{revtex4-1}
\usepackage{graphicx}
\usepackage{amsmath}
\usepackage{bm}
\usepackage{amssymb}
\usepackage{color,soul}
\usepackage{ulem}  
\usepackage{amsfonts}%
\usepackage[caption=false]{subfig}
\usepackage{comment}
\usepackage{color,soul}
\usepackage[dvipsnames]{xcolor}
\usepackage[capitalize]{cleveref}
\usepackage{float}
\usepackage{tabularx} 
\usepackage{array}   
\newcolumntype{L}{>{$}l<{$}} 
\newcolumntype{R}{>{$}r<{$}} 
\newcolumntype{C}{>{$}c<{$}} 
\usepackage{comment}
\usepackage[capitalize]{cleveref}

\newcommand{\itsection}[1]{\textit{#1.}---}

\newcommand\thao[1]{\textcolor{blue}{(TN: #1)}}
\newcommand\ky[1]{\textcolor{red}{(KY: #1)}}

\newcommand\blue[1]{{\color{blue}#1}}

\newcommand\mcl{$M\rm{Cl_2}$}
\newcommand\vcl{VCl$_2$}
\newcommand\mncl{MnCl$_2$}
\newcommand\nicl{NiCl$_2$}

\begin{document}

\title{
Goodenough-Kanamori-Anderson rules in 2D  magnet: \\
A chemical trend in $\bm{M}$Cl$_2$ 
with $\bm M$=V, Mn, and Ni 
}


\author{Thi Phuong Thao Nguyen}
\author{Kunihiko Yamauchi}
\affiliation{%
Department of Precision Engineering, Graduate School of Engineering, Osaka University, 2-1 Yamadaoka, Suita, Osaka 565-0871, Japan
}%
\affiliation{%
 Center for Spintronics Research Network, Osaka University, 1-3 Machikaneyama, Toyonaka, Osaka 560-8531, Japan
}%

\date{\today}

\begin{abstract}

Density-functional-theory calculations were performed 
to investigate the magnetism in  a series of triangular-lattice monolayer \mcl\ ($M$=V, Mn, and Ni).  
The magnetic stability manifests a distinct chemical trend; VCl$_2$ and MnCl$_2$ show the antiferromagnetic ground states and  NiCl$_2$ shows the ferromagnetic ground state. 
The microscopic mechanism behind the magnetic interaction is explained by the so-called Goodenough-Kanamori-Anderson rules and by the  virtual-hopping process
through the 
hopping integrals between the 3$d$-orbital maximally localized Wannier functions. 
Our result highlights the role of the direct exchange interaction and the superexchange interaction 
in the magnetic stabilization 
in two-dimensional magnets. 
\end{abstract}

\maketitle

\section{Introduction}

\begin{table}[b]
    \centering
    \setlength{\tabcolsep}{2pt}
     \resizebox{65mm}{!}{%
    \begin{tabular}{r| c c c c r }
    \hline
    \hline
  \multicolumn{1}{c}{} & \multicolumn{3}{c}{Bulk (experiment)}\\ 
    \hline
 &   & 
 $T_{\rm C}$ & \begin{tabular}{c}Magnetic \\ order\end{tabular} & \begin{tabular}{c}In-plane \\ order \end{tabular}  \\
    \hline
    VCl$_2$~\cite{Kadowaki1986_vcl2}  &  & 35.8 & ncl  &  120$^{\circ}$-AFM  \\
    MnCl$_2$\cite{Murray1962_MnCl2}   &   & 2.0, 1.8  & unclear &  unclear   \\ 
    NiCl$_2$\cite{Ferrari1962_nicl2} &  & 52 & col  &  FM    \\ 
    \hline
    \hline
    \end{tabular}
    }
    \caption{ 
    Summary of experimentally reported magnetic property for bulk $M$Cl$_2$  compounds. 
    $T_{\rm C}$ (K)is the ordering  temperature of collinear (col) or noncollinear (ncl) AFM and FM orderings. In-plane order shows the arrangement of the magnetic moments in a single atomic layer.}
\label{table:summary}
\end{table}

\begin{figure}[t]
    \centering
    \includegraphics[width=80mm, angle=0]
    {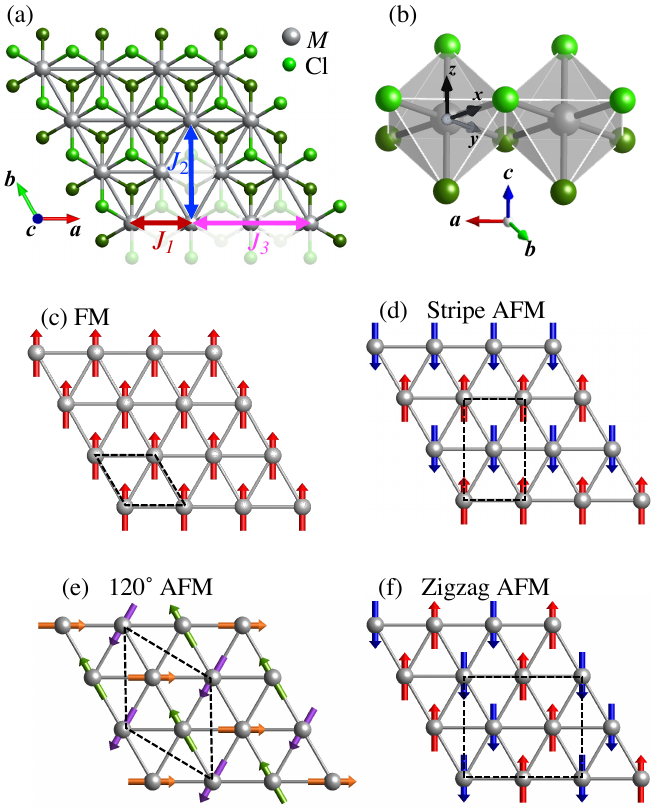}
    \caption{(a) Crystal structure of 
    monolayer $M$Cl$_2$ in a triangular lattice (corresponding to 1T phase in bulk). Dark and light green balls represent Cl atoms in the lower and upper layers, respectively,  
    while grey balls represent $M$ atoms. 
The double-headed arrow indicates the in-plane exchange coupling $J_{ij}$. (b) Neighboring $M$Cl$_6$ octahedra sharing the edge. 
The local coordinate system $\{xyz\}$ in an octahedron and the hexagonal lattice vectors $\{abc\}$  are shown by black and colorful arrows, respectively. 
(c-f) Magnetic configurations considered in this study. The black dashed line indicates the magnetic supercells used in the calculations. 
}
    \label{fig:mcl2_afm_config} 
\end{figure}

Recent discoveries of two-dimensional (2D) magnetic van der Waals (vdW) materials have ignited the research boom in exploring further 2D magnetic materials to take advantage of their atomically thin thickness and to examine the novel spintronics applications.~\cite{Gibertini_2019review, Wang_2022review, McGuire2017_review}
So far, 2D ferromagnetism has been reported in atomically thin layers  
of Cr$_2$Ge$_2$Te$_3$, CrI$_3$ and VI$_3$ in which the transition-metal ion forms a honeycomb structure.~\cite{Gong2017_cr2ge2te3, cri3, cri3_doping, vi3_bulk2_insu, thao2021} 
%
On the other hand, 2D antiferromagnetism has been reported in NiPS$_3$ and MnPS$_3$, \cite{Kim2019_NiPS3, Kim_20192DMater}. 
They are candidate materials for antiferromagnetic (AFM) spintronics applications possibly used in    
low-energy-consuming  and high-speed devices.  \cite{Baltz2018_AFMspintronics, Sharidya2021_review2DAFM} 
In principle, an AFM interaction in the 2D triangular lattice causes magnetic frustration and often results in non-collinear magnetic orderings such as 120$^\circ$ AFM configuration shown in Fig.~\ref{fig:mcl2_afm_config} (e).~\cite{ Kulish2017, luo2020, Yekta2021_MX2_MX3, Zhang_2022APL_FeO2} 
When a strong spin-orbit coupling (SOC) is involved in the system, a {\it chiral} spin spiral state can be stabilized through an anisotropic magnetic exchange or  Dzyaloshinskii-Moriya interaction. 
It has been reported that monolayer NiI$_2$ and NiBr$_2$ show fascinating 2D multiferroic behavior;  the ferroelectric polarization is induced by the chiral magnetic texture owing to  the strong SOC at the heavy halogen atoms. \cite{Danila2020_NiI2, Song_Silvial2022_NiI2, Fumega2022}   
While the role of the SOC has been well discussed in previous studies, the basic discussion on magnetic frustration based on the Heisenberg magnetic interactions has not been particularly focused.~\cite{Prayitno2019_MnCl2, Kira2022, Wang2022_GKA_2Dmagnets}  


%
Along this context, to have a deeper understanding of the AFM interaction and the magnetic frustration  in the monolayer triangular lattice,  
in this work, we investigate a chemical trend of magnetism in 
monolayer transition-metal dihalides \mcl\ ($M=$ V, Mn, Ni) 
in connection with their electronic state. 
There are two reasons for the choice of these target materials: 1) Since they are composed of light elements, we can ignore the SOC effect and focus on the Heisenberg interactions; 2) in their bulk form, a rich variety of magnetic ground states has been experimentally observed, and it helps us to understand the chemical trend of the magnetism with respect to the different $3d$ electron filling   
(see Table~\ref{table:summary}). 
While it has been reported that \vcl\ shows 120$^\circ$ AFM ordering, and \nicl\ shows FM ordering at low temperatures~\cite{kadowaki1987experimental, Ferrari1962_nicl2}, the magnetic ground state 
in \mncl\ is still unclear~\cite{Murray1962_MnCl2}; recent studies suggested the possibility of its helimagnetic ground state.~\cite{Ishii_JPSJ2019, Wiesler_PRB1997}   
In the rest of this paper, first, we will discuss the magnetic stability and the exchange coupling in the monolayer halides. 
Secondly, using the virtual hopping interpretation for magnetic stability, we will discuss the microscopic mechanism behind the magnetic interaction.



\section{Methodology}
Transition-metal dihalides $MX_2$  
form a triangular lattice in which the cation $M$ is surrounded by six  ligand $X$ anions  in octahedron coordination, as illustrated in Fig.~\ref{fig:mcl2_afm_config} (a). 
In the bulk form, \vcl\  crystallize in the trigonal CdI$_2$ structure ($P\bar{3}m1$ space group) and \mncl\ and \nicl\ crystallize in the rhombohedral CdCl$_2$ structure ($R\bar{3}m$ space group). These two structure types have different stackings along the \textit{c}-axis, whereas they have the same in-plane atomic structure. Individual layers of dihalides can be extracted from the bulk using exfoliation techniques. 
To calculate the electronic state in the monolayer $M$Cl$_2$, we set up the slab model containing one monolayer and 18 \AA\ length of vacuum region in a periodic supercell. 

Density-functional theory (DFT) calculations were performed using VASP \cite{vasp, paw} code with  generalized gradient approximation (GGA)~\cite{gga}. Hubbard 
$U$ potential, $U$ = 1.8 eV and $J_{\rm H}$ = 0.8 eV for V, Mn, and Ni $3d$ orbital state, was employed within the Liechtenstein approach.~\cite{Liechtenstein, Kira2022}
The atomic positions and the lattice parameters were optimized in  FM configuration until forces acting on atoms were smaller than 0.1 eV/\AA\ (the results are shown in Table.~\ref{table:structure}). 
To evaluate the magnetic stability and to extract the Heisenberg exchange coupling constants, $J_{ij}$, we considered four different spin order arrangements, that is, ferromagnetic (FM), stripe-type AFM, 120$^\circ$ frustrated AFM, and zigzag-type AFM configurations, shown in Fig. 1 (c-f). 
The minimum size of supercells was used for each AFM configuration --- containing two, 
three, and four f.u. for  stripe AFM, 120$^\circ$ AFM, and zigzag AFM configurations, respectively --- to reduce the computational cost.  
While the $k$-point mesh was set to be
18 $\times$ 18 $\times$ 1 for the unit cell, we used 
$18\times10\times1$, 
$10\times10\times1$,  
and $10\times9\times1$, for these AFM supercells, respectively.  
The difference of total energy, $\Delta E=E_{\rm AFM} - E_{\rm FM}$,  was calculated between AFM and FM configurations in the same supercell and with the same $k$-point mesh. 
By taking the difference, we can eliminate the energy terms that depend on the supercell shape and extract the energy purely related to the magnetic stability. 
%

After the self-consistent calculation in FM configuration converged, the Bloch functions relevant to five $M$-$d$ bands and six Cl-$p$ bands (for each spin state) were projected onto the maximally localized Wannier functions (MLWFs) by using the Wannier90 tool 
to evaluate the hopping parameters 
between the $M$-$3d$ orbital states.~\cite{wannier90, Marzari2012, Pizzi_2020} 
%
%
The obtained hopping parameter $t$ implicitly contains both the direct $d-d$ hopping and the indirect $d-p-d$ hopping processes since the Wannier functions have delocalized character reflecting the hybridization between $M$-$3d$ orbital states and ligand Cl-$p$ orbital states. 

\section{Results and discussions}

\subsection{Electronic state and crystal-field splitting}

\begin{figure}[t]
    \center
\includegraphics[width=70mm, angle=0]{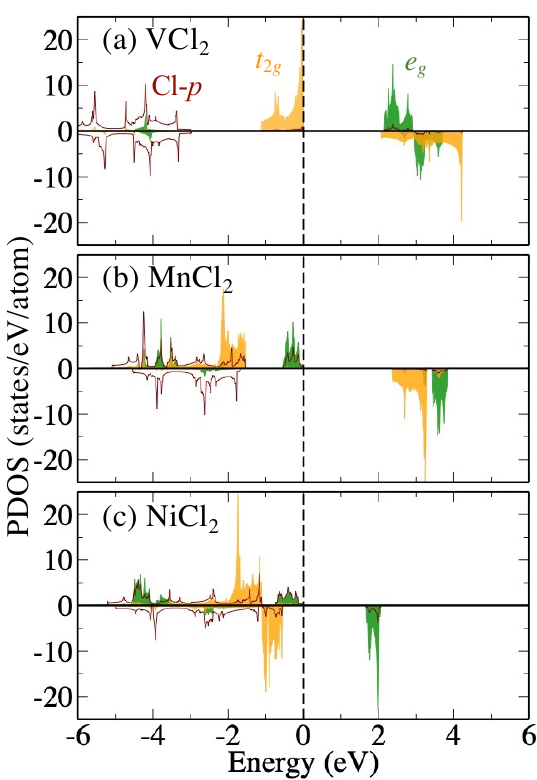}
    \caption{
   DOS projected onto transition-metal 3$d$ and Cl-2$p$ state with $O_h$ crystal-field states via MLWFs. 
    Green (orange) filled-curve represents $e_g$ ($t_{2g}$) states. The brick red curve represents for $p$ state.
    Fermi level is set at energy origin. 
\label{fig:DOS}   }
\end{figure}
Figure \ref{fig:DOS} shows the calculated density of states (DOS). 
Since the octahedral crystalline electric field splits the $d$ orbital states into the two-fold $e_g$ and three-fold $t_{2g}$ states, 
%
all the materials show the insulating  state with 
the band gap, 2.0 eV, 2.36 eV and 1.66 eV for VCl$_2$, MnCl$_2$ and NiCl$_2$, respectively, which is in agreement with previous studies.~\cite{Kulish2017, botana2019electronic, Kira2022}  
In \vcl, the V$^{2+}$ ($d^3$) ion exhibits the high-spin ($S=3/2$) state 
with the empty $e_g^{\uparrow}$ and occupied $t_{2g}^{\uparrow}$ states in the majority spin state. 
There is no significant splitting between $e_g^{\downarrow}$ and $t_{2g}^{\downarrow}$ states in the minority spin state due to  the weak hybridization of the $d^{\downarrow}$ states in higher energy with the Cl-$p$ states located below the Fermi energy. In MnCl$_2$, the Mn$^{2+}$ ($d^5$)  ion exhibits the  high-spin ($S=5/2$) state with the fully occupied (unoccupied) majority-spin (minority-spin) state. 
In NiCl$_2$ the Ni$^{2+}$ ($d^8$) ion exhibits the high-spin ($S=1$) state with the fully occupied majority-spin state and the occupied $t_{2g}^{\downarrow}$ state and the empty $e_{g}^{\downarrow}$ state in the minority-spin channel. 


\begin{table}[t]
\centering\def\arraystretch{1.1} \setlength{\tabcolsep}{3pt}
 \resizebox{80mm}{!}{%
 \begin{tabular}{c | c c c c c c } 
 \hline
 \hline
 & $a$  & $\theta$ & $M$-Cl & NN$^{\rm 1st}$ & NN$^{\rm 2nd}$ & NN$^{\rm 3rd}$ \\
  \hline
VCl$_2$ &  3.54 &  91.07$^{\circ}$ & 2.48 & 3.54 & 6.13 & 7.08  \\
MnCl$_2$ &   3.70 &  93.16$^{\circ}$ & 2.55 & 3.70 & 6.41 & 7.40  \\
  NiCl$_2$ & 3.49 & 93.27$^{\circ}$  & 2.40 & 3.49 & 6.04 & 6.98 \\
  \hline
  \hline
  \end{tabular}
  }
 \caption{ 
Optimized structural parameters: in-plane lattice constant $a$ ({\AA}); $M$-Cl-$M$ angle $\theta$ ($^{\circ}$), $M$-Cl bond lengths ({\AA}); distance ({\AA}) between first (NN$^{\rm 1st}$), second (NN$^{\rm 2nd}$), and third (NN$^{\rm 3rd}$) nearest neighbor   transition-metal-ionic sites. 
\label{table:structure}}
\end{table}

\subsection{Magnetic stability and exchange interaction}
The transition-metal spin momenta were calculated as 2.63 $\mu_{\rm B}$, 4.51 $\mu_{\rm B}$ and 1.51 $\mu_{\rm B}$ 
for VCl$_2$, MnCl$_2$ and NiCl$_2$, respectively, comparable to their nominal spin sizes, $S$= 3, 5, 2 $\mu_{\rm B}$, respectively. 
%
Table \ref{table:magnetic} shows the evaluated magnetic stability for the monolayers. VCl$_2$ strongly favors 120$^\circ$ AFM spin configuration, being consistent with  
a neutron diffraction study for bulk VCl$_2$ reporting the in-plane 120$^\circ$ AFM spin configuration (see also \cref{table:summary}).\cite{Kadowaki1986_vcl2} 
In MnCl$_2$, although 120$^\circ$ AFM order is slightly preferred, three AFM spin configurations are very close in the energy.  
A neutron diffraction study reported that 
there are two magnetic phase transitions at $T_{\rm N}$=1.81 and 1.96 K, but the spin configuration at the ground state has not been clarified.\cite{Murray1962_MnCl2}    
Making a keen contrast, NiCl$_2$ shows a strong preference for FM ordering. This is consistent with previous experimental findings of the in-plane FM configuration with out-of-plane AFM ordering. 
\cite{Busey1952_NiCl2, Katsumata1973_NiCl2, Ferrari1962_nicl2}

\begin{figure}[t]
    \center
\includegraphics[width=70mm, angle=0]{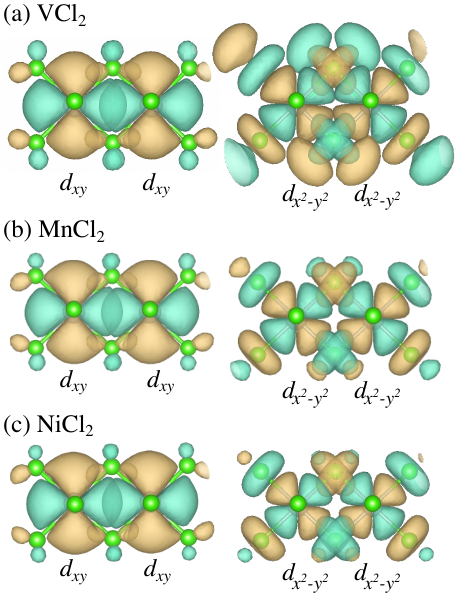}
    \caption{
    The MLWFs relevant to 
    the two dominant nearest-neighbor hoppings between $d_{xy}$ and $d_{xy}$,  and between $d_{x^2-y^2}$ and $d_{x^2-y^2}$ orbitals.
   Isosurface level was set at $\pm 0.2$ $a_0^{-3/2}$, where $a_0$ is the
Bohr radius. 
\label{fig:MLWFs}   }
\end{figure}

To understand the magnetic stability, we evaluated the magnetic exchange interactions ($J_{ij}$) between $M$ atoms  by fitting the total energy differences to the Heisenberg Hamiltonian:  
\begin{equation}
    \label{eq:Heisenberg}
    H=\sum_{<ij>} J_{ij}s_i\cdot s_j
\end{equation}
where $s_i$ and $s_j$ are the spin unit vectors at the site $i$ and site $j$.  When $J<0$ ($J>0$), a parallel spin (anti-parallel spin) configuration is preferred. In the triangular lattice monolayer, we consider $J_1$, $J_2$, and $J_3$ exchange couplings, corresponding to the first, second, and third nearest neighbor  interactions, as illustrated in~\cref{fig:mcl2_afm_config} (a). 
In order to derive these parameters, the total energy of four magnetic configurations in ~\cref{fig:mcl2_afm_config} are mapped onto the Heisenberg Hamiltonian. Equation~\ref{eq:Heisenberg} for the AFM configurations 
can be written as 
\begin{eqnarray}
     \label{eq:Energy mapping}
    E_{\rm stripe}=-2J_1-2J_2+6J_3,  \nonumber \\
    E_{120^{\circ}}=-9/2J_1+9J_2-9/2J_3, \nonumber \\
    E_{\rm zigzag}=-4J_1+4J_2-4J_3, 
\end{eqnarray}
 per each magnetic supercell. Then we calculated the difference between AFM and FM total energy as $\Delta E = E_{\rm{AFM}} - E_{\rm{FM}}$ for each magnetic supercell and solved the simultaneous equations for $J_1$, $J_2$, and $J_3$.     
The FM total energy is defined as $E_{\rm FM} = 3n (J_1+J_2+J_3) \nonumber $ where $n$ = 2, 3, 4  denotes the size of the magnetic supercell.

The ferromagnetic and antiferromagnetic transition temperatures ($T_{\rm C}$ and $T_{\rm N}$) were evaluated within the mean-field approximation~\cite{Liech1987JMMM, Okuyama_KY} as follows:
\begin{eqnarray}
    \label{eq:mean-field}
    T_{\rm C}=-1/3k_{\rm B} (3J_1+3J_2+3J_3),\\
    T_{\rm N}=-1/3k_{\rm B}(-3/2J_1+ 3J_2-3/2J_3),
\end{eqnarray}
where $k_{\rm B}$ is the Boltzmann constant.

 \begin{table}
\centering\def\arraystretch{1.1}
\setlength{\tabcolsep}{7pt}
 \resizebox{75mm}{!}{%
 \begin{tabular}{l |cccc}
 \hline
 \hline
  $\Delta E$ & Stripe & 120$^{\circ}$ & Zigzag & FM \\
  \hline
  VCl$_2$ & $-78.41$ & ${\bf -85.53}$ & $-75.07$ & 0.00\\
  MnCl$_2$ & $-8.82$ & ${\bf -10.23}$ & $-9.94$  & 0.00 \\
  NiCl$_2$ &  +19.82 & +17.57 & +13.09 & ${\bf 0.00}$ \\
  \hline
  \end{tabular}
  }
  \resizebox{75mm}{!}{%
   \begin{tabular}{l |ccc| c}
 \hline
 $J_{ij}$ & $J_1$ & $J_2$ & $J_3$ & $T_{\rm C}$, $T_{\rm N}$ \\
 \hline
  VCl$_2$ & +20.08 & $-0.48$ & $-1.07$ & 115.8 \\
  MnCl$_2$ & +1.78 & +0.43  & +0.49 & 8.3\\
  NiCl$_2$ & $-6.22$ & +1.26 & +2.31 & 30.7\\
  \hline
  \hline
  \end{tabular}
  }
 \caption{ Top: Calculated total energy  $\Delta E$ (meV/f.u) for each magnetic configuration with respect to FM configuration; 
 the lowest energy is highlighted. 
 Bottom: Exchange coupling constants 
 $J_1$, $J_2$, $J_3$ (meV) and 
the critical temperature $T_{\rm C}$ and $T_{\rm N}$ (K). 
}
\label{table:magnetic}
\end{table}

The calculated magnetic exchange coupling constants are tabulated   in~\cref{table:magnetic}. The first nearest interaction $J_1$ is the  dominant interaction that leads to the 120$^\circ$-AFM ground state in VCl$_2$ and MnCl$_2$,  and FM ground state in NiCl$_2$. 
While the estimated critical temperatures are overestimated owing to the mean-field approximation, we can still qualitatively discuss the magnetic stability along the chemical trend.  
In MnCl$_2$, $J_1$, $J_2$, and $J_3$ show all small positive values and lead to 
the competing stability  between several AFM configurations. 
This may be responsible for the theoretically predicted spin-spiral ground state.\cite{Ishii_JPSJ2019} 
For all three chlorides, the longer-range interaction $J_2$ is much smaller than $J_1$, while $J_3$ is found to be larger than $J_2$. This trend ($|J_1|>> |J_3| > |J_2|$) is  consistent with a previous theoretical work~\cite{Kira2022} 
 and characteristic of the triangular lattice;  
the edge-sharing $M$Cl$_6$ octahedral geometry allows both direct ($d-d$ or $d-d-d$) hoppings and indirect ($d-p-d$ or $d-p-d-p-d$) hoppings in $J_1$ and $J_3$ interactions while only the indirect  ($d-p-d-p-d$) hopping is possible for $J_2$ interaction.




To understand the $J_1$ character for these chlorides, we apply the 
Goodenough-Kanamori-Anderson (GKA) rules that commonly explain the magnetic interaction in transition-metal oxides 
to the superexchange interaction in the 90$^{\circ}$ angle $M$-Cl-$M$ bond. 
According to the GKA rule, ---the third rule made for 90$^{\circ}$ bonds---, 
1) $e_g-e_g$ exchange is ferromagnetic and weak, 2) direct $t_{2g}-t_{2g}$ interaction can be antiferromagnetic and 3) indirect $t_{2g}-p-t_{2g}$ and $t_{2g}-p-e_{g}$ interactions can be either strong antiferromagnetic or  (weak) ferromagnetic  dependent on the orbital occupation.~\cite{Goodenough1955, Kanamori1958, Anderson1950, khomskii_book} 
%
%
Therefore, we can deduce a consistent conclusion.  
For \vcl, $t_{2g}^{3\uparrow}-p-t_{2g}^{0\downarrow}$ exchange may result in the AFM interaction.
For \mncl, $e_{g}^{2\uparrow}-p-t_{2g}^{0\downarrow}$ exchange may cause FM interaction, but also  $t_{2g}^{3\uparrow}-p-t_{2g}^{0\downarrow}$ exchange may counteract, and thus the net interaction may become weak. 
For \nicl, $e_{g}^{2\uparrow}-p-t_{2g}^{0\downarrow}$ may result in FM interaction. 
%
This is also consistent with a discussion made by Kanamori on $90^{\circ}$ superexchange interactions between V$^{2+}$ sites ($d^3-d^3$) and Ni$^{2+}$ sites ($d^8-d^8$) concluding, resulting in AFM and FM couplings, respectively.~\cite{Kanamori1958} 
%
%

In the current study, we will take a step further by presenting a quantitative analysis of  the $d-d$ hopping integrals. 
In Mott insulators, the direct exchange coupling is approximated as 
$J_{\rm } \sim \sum_{m,n} 2{t_{mn}}^2/\Delta E_{mn}$
with $d-d$ hopping integral $t_{mn}$ between occupied $m$ and unoccupied $n$ orbital states  at neighboring transition-metal sites and the difference of those energy levels $\Delta E_{mn}$  (in the case of two orbital state model, it corresponds to  on-site Coulomb repulsion $U_{\rm }$ that separates the filled and the unfilled energy levels); this process is known as  \textit{``virtual hopping''}~\cite{khomskii_book}. 
If  the hopping is stronger between orbital states that belong to the same spin states than that for opposite spin states, it leads to  a parallel-spin configuration; on the other hand, if hopping between orbitals that belong to the opposite spin states is stronger, it leads to the anti-parallel spin configuration.  
Hereinafter, we assume that the hopping integral is not significantly modified by a change in the spin configurations and in the energy level,  
and therefore, we evaluate the orbital dependency of the hopping integral calculated in FM state. 

\begin{table}[t]
    \centering\def\arraystretch{1.1}
    \setlength{\tabcolsep}{3.5pt}
\resizebox{65mm}{!}{%
    \begin{tabular}{ R|RRRRR }
        \multicolumn{1}{C}{} & \multicolumn{5}{C}{\text{ }} \\
 {\rm VCl_2} &d_{xy}  &  d_{yz}  & d_{zx} & d_{x^2\text{-}y^2} &  d_{z^2}  \\
 \hline
d_{xy}  & \mathbf{-250} &12 &13 &4 &-33\\
d_{yz}  & 12 &68 &-38 &-12 &-2\\
d_{zx} & 13 &-38 &72 &11 &-2\\
d_{x^2\text{-}y^2} & 4 &-12 &11 &\mathbf{-162} &2\\
d_{z^2} & -33 &-2 &-2 &2 &-7\\
       \multicolumn{1}{C}{} & \multicolumn{5}{C}{\text{ }} \\
 {\rm MnCl_2} &d_{xy}  &  d_{yz}  & d_{zx} & d_{x^2\text{-}y^2} &  d_{z^2} \\
    \hline
d_{xy}  & \mathbf{-81} &7 &7 &1 &0\\
d_{yz}  & 7 &20 &-9 &-7 &-1\\
d_{zx} & 7 &-9 &21 &6 &-1\\
d_{x^2\text{-}y^2} & 1 &-7 &6 &\mathbf{-67} &0\\
d_{z^2} & 0 &-1 &-1 &0 &7\\
        \multicolumn{1}{C}{} & \multicolumn{5}{C}{\text{ }} \\
 {\rm NiCl_2} &d_{xy}  &  d_{yz}  & d_{zx} & d_{x^2\text{-}y^2} &  d_{z^2} \\
    \hline
d_{xy}  &\mathbf{-77} &7 &7& 0 &4\\
d_{yz}  &7 &20 &-9 &-7 &0\\
d_{zx} &7 &-9 &21 &6 &0\\
d_{x^2\text{-}y^2} &0 &-7 &6 &\mathbf{-69} &0\\
d_{z^2} &4 &0 &0 &0 &13\\
  \end{tabular}
  }
    \caption{
    \label{tab:hopping}
    Hopping integrals $t_{dd}$ (meV) calculated by MLWF basis set of the nearest-neighbor $M$-$d$ orbital states. The dominant elements are highlighted. }
\end{table}

The calculated $d-d$ hopping integrals $t_{dd}$ 
between the first nearest neighbor  $M$ atomic cites are shown in \cref{tab:hopping}. 
A common feature among those chlorides is found that 
 $t_{xy, xy}$  and $t_{x^2-y^2, {x^2-y^2}}$ 
 dominate over the $d-d$ hopping integrals.  
As is typical in the edge-sharing octahedra, neighboring $d_{xy}$ orbitals can directly interact through the orbital lobes pointing toward each other, resulting in the largest hopping integral. 
On the other hand, $t_{x^2-y^2, {x^2-y^2}}$ 
shows rather indirect hopping characteristic. The tails of the neighboring Wannier functions show a large overlap in the vicinity of Cl sites. This leads to $d-p-p-d$ superexchange interaction.    
Both $t_{xy, xy}$ and $t_{x^2-y^2, {x^2-y^2}}$ hopping integrals manifest an obvious chemical trend --- they decrease  from V to Mn to Ni. 
This is explained by the incomplete screening of core potential.~\cite{HaverkortAndersen.PRB2012} 
When one electron and one proton are added to an atom, the electron incompletely screens the attractive potential from the proton seen by another valence electron. 
It makes  the one-electron potential become deeper and deeper as the valence electrons increase. 
Therefore, Ni-$d$ orbital state is more localized than V-$d$ orbital state, and the hopping is suppressed (compare the spreading of the Wannier functions  in Fig.~\ref{fig:MLWFs}). 

\begin{figure}
    \centering
\includegraphics[width=80mm, angle=0]{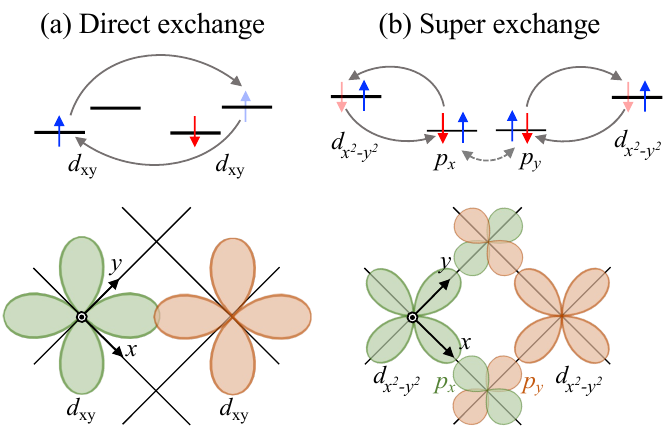}
    \caption{
Schematic pictures of (a) direct exchange and (b) superexchange interactions between the nearest-neighbor $M$-$3d$ orbitals. Direct exchange stabilizes an anti-parallel spin configuration due to the virtual hopping process. Super exchange stabilizes a parallel spin configuration due to the virtual hopping process together with the Hund's coupling at Cl site.   
}
    \label{fig:schematic_hopping}
\end{figure}

The direct $d_{xy}-d_{xy}$ hopping leads to an anti-parallel spin configuration by the direct exchange process if the electron occupying the  majority-spin state can hop to the empty minority-spin state in a neighboring atom (see Fig.~\ref{fig:schematic_hopping} (a)). This causes anti-parallel-spin coupling in VI$_2$ where $d_{xy}$ state is occupied only in one spin channel. 
On the other hand, the indirect $t_{x^2-y^2, {x^2-y^2}}$ hopping leads to a parallel spin configuration by the superexchange process mediated by the Cl-$p$ Hund's coupling that favors parallel spin configuration at Cl site if  Cl-$p$ electron can hop onto the partially unoccupied $d_{x^2-y^2}$ orbital state (see Fig.~\ref{fig:schematic_hopping} (b)). 
This causes parallel-spin coupling in NiCl$_2$, where $d_{x^2-y^2}$ state is occupied only in one spin channel. 
In MnCl$_2$, both the direct and the superexchange processes are possible to take place; their competition ends up with its weak magnetic interaction. 

\section{Conclusions}
By means of density-functional-theory calculations, 
we investigated the magnetism  in monolayer $M$Cl$_2$. Our result shows good agreement with the experimental reports; VCl$_2$ favors the AFM configuration,  NiCl$_2$ favors the FM configuration, and MnCl$_2$ shows the weak magnetic interaction.  
Considering the chemical trend, we confirmed that the Goodenough-Kanamori-Anderson rules still stand in the monolayer transition-metal chlorides. On top of it, it should be  emphasized that  $d_{xy}$ and $d_{x^2-y^2}$ orbital states play an important role to determine the exchange interaction in the 2D magnets. 
This finding may provide insights into the  magnetism, typically associated with the frustrated magnetism such as helimagnetism and skyrmion in recently developed 2D materials. 

\begin{acknowledgments}
This work was supported  by JST-CREST (Grant No. JPMJCR18T1). The computation in this work has been done using the facilities of the Supercomputer Center, the Institute for Solid State Physics, the University of Tokyo and Supercomputing System MASAMUNE-IMR in the Center for Computational Materials Science, Institute for Materials Research, Tohoku University (Project No. 20K0045). The crystallographic figure was generated using the VESTA program~\cite{vesta3}.
\end{acknowledgments}

\appendix*

\crefalias{section}{appsec}
\section{Electronic band structure  }
\label{sec:appendix}
\renewcommand{\thefigure}{A-\arabic{figure}}
\setcounter{figure}{0}
\begin{figure}[t]
    \center
\includegraphics[width=60mm, angle=0]{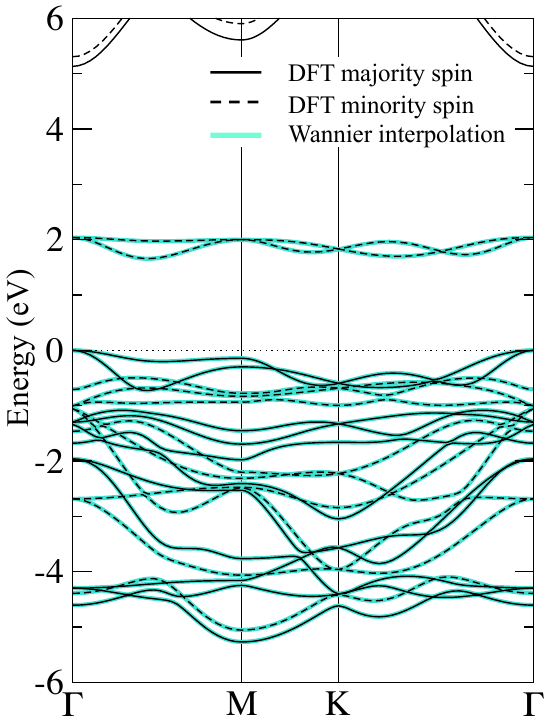}
    \caption{ Energy bandstructure obtained from the DFT calculations (black solid and dashed lines) and from the MLWF interpolation (cyan lines) of monolayer NiCl$_2$  along the high-symmetry lines. The Fermi energy is set at zero. 
   }
   \label{fig:band}
\end{figure}

Figure \ref{fig:band} shows the calculated bandstructure of monolayer \nicl\ fitted by the MLWF bands.

\bibliography{mybibitem.bib}
\bibliographystyle{apsrev4-1}  

\end{document}